\documentclass[twocolumn,showpacs,preprintnumbers,amsmath,amssymb,pra,floatfix]{revtex4}
\usepackage{graphicx}% Include figure files
\usepackage{dcolumn}% Align table columns on decimal point
\usepackage{bm}% bold math

\begin{document}

%\preprint{APS/123-QED}

\title{Ferromagnetic Resonance in Spinor Dipolar Bose--Einstein Condensates}

\author{Masashi Yasunaga}
% \altaffiliation[Also at ]{Department of Physics, Osaka City University}%Lines break automatically or can be forced with \\
\author{Makoto Tsubota}%
 \email{tsubota@sci.osaka-cu.ac.jp}
\affiliation{
Department of Physics, Osaka City University, Sumiyoshi-ku, Osaka 558-8585, Japan
}%
\date{\today}% It is always \today, today,
             %  but any date may be explicitly specified

\begin{abstract}

We used the Gross--Pitaevskii equations to investigate ferromagnetic resonance in spin-1 Bose--Einstein condensates with a magnetic dipole-dipole interaction. By introducing the dipole interaction, we obtained equations similar to the Kittel equations used to represent ferromagnetic resonance in condensed matter physics. These equations indicated that the ferromagnetic resonance originated from dipolar interaction, and that the resonance frequency depended upon the shape of the condensate. Furthermore, spin currents driven by spin diffusions are characteristic of this system. 
		   
\end{abstract}

\pacs{03.75.Mn, 03.75.Nt }% PACS, the Physics and Astronomy
                             % Classification Scheme.
%\keywords{Suggested keywords}%Use showkeys class option if keyword
                              %display desired
\maketitle

\section{INTRODUCTION}\label{sec:introduction}

Magnetic resonance (MR) as a physical concept has been applied in various fields, enabling physical, chemical, and medical experiments to obtain information on nuclear spin and electron spin systems. The concept has also provided valuable information to help understand the unknown structures of many condensed matter systems \cite{Slichter}. 

The use of MR in the study of ferromagnets, {\it e.g.} Nickel, Cobalt, and Iron, began in the 1940s. Griffiths observed that the Land\'e's $g$-factor of electrons in ferromagnets was far from the well known value, 2 \cite{Griffiths1946}.  In order to understand these anomalous results, Kittel theoretically introduced a demagnetizing field into the equation representing the motion of the magnetization ${\bf M} = (M_x, M_y, M_z)$, obtaining an equation valid in an external magnetic field $H_0\hat{{\bf z}}$, with $M_{z0} = H_0/N_z$ and demagnetizing fields \cite{Kittel1947_8}, thereby obtaining the Kittel equation,
\begin{equation}\label{eq:Kittel}
\frac{d {\bf M}}{dt} = \gamma _n [ {\bf M} \times {\bf H}].
\end{equation}
Here, $\gamma _n$ is the nuclear gyromagnetic ratio, and ${\bf H} = (-N_xM_x, -N_yM_y, H_0-N_zM_z) $ is given by the demagnetizing factors $N_i$. By linearizing the magnetization ${\bf M} = {\bf M}_0 + \delta {\bf M}$ from the stationary magnetization ${\bf M}_0 = M_{z0}\hat{{\bf z}}$, Kittel obtained a precession of the magnetization and a precessing frequency, {\it i.e.} resonance frequency,
\begin{equation}\label{eq:wk}
\omega^2 = \gamma _n^2 \{H_0+(N_y-N_z)M_{z0}\}\{H_0+(N_x-N_z)M_{z0}\},
\end{equation}
 which explained the anomalous $g$-factor. Furthermore, he found that the resonance frequency depends on the shape of a ferromagnet because $N_i$ depends on the shape \cite{Kittel1947_8}. Thus, ferromagnetic resonance (FMR) was established, and the work enabled numerous additional studies \cite{Kittel2005}.

MR also plays an important role in quantum condensate systems. In superfluid $^3$He, the dynamics of the spin vector and the $d$-vector are represented by the Leggett equation, which couples these vectors through magnetic dipole-dipole interactions \cite{Leggett1975}. The equation also shows not only an MR typical of condensed matter, but also a new MR that cannot be described using the equations of motion for general paramagnets and ferromagnets. This MR was used to find  $A$ and $B$ phases \cite{Vollhard}. Parallel ringing, which is an oscillation of longitudinal magnetization, was also observed \cite{Webb1974}.

Since the discovery of atomic Bose--Einstein condensates (BECs) \cite{Davis1995, Anderson1995}, BECs have been studied in optics and atomic and condensed matter physics. We have introduced MR into BECs to realize magnetic resonance imaging, a popular method of nondestructive testing. Spinor BECs are expected to be suitable for MR, since they have not only internal degrees of freedom but also magnetic properties. In particular, we are interested in magnetic dipole-dipole interactions (MDDI) in spinor BECs, which have been actively studied. The interaction between spins has a characteristic symmetry of rotation and spin, which is expected to result in a new quantum phase \cite{Yi2004, Yuki2006a, Makela2007} and Einstein--de Haas effects \cite{Yuki2006b}. Experimentally, Griesmaier {\it et al.} realized spinor dipolar condensates using $^{52}$Cr atoms, which have a larger magnetic moment than alkali atoms \cite{Griesmaier2005}. The shape of the condensates clearly represented the anisotropy of the interaction \cite{Stuhler2006, Lahaye2007}. Thus, MDDI has opened new areas of spinor condensate research.

As an introduction to MR in BECs, we numerically studied spin echo in dipolar BECs with spin-1 \cite{Yasunaga2008}. The spin echo is a typical phenomenon of MR, discovered by Hahn \cite{Hahn1950} and developed by Carr and Purcell \cite{Carr1954}. Previously, we calculated the transition from Rabi oscillations to internal Josephson oscillations in spinor condensates \cite{Yasunaga2009}. In this paper, we consider MDDI in spin-1 BECs, examining FMR by analyzing the Gross--Pitaevskii (GP) equations. 

In section \ref{sec:formulation}, we derive Kittel-like equations from the GP equations, and analyze them. In section \ref{sec:FMRunderSMA}, using a single-mode approximation, we derive Kittel equations from the Kittel-like equations. The MDDI of the Kittel equations is considered as the origin of the demagnetizing field, which is phenomenologically introduced in Eq. \eqref{eq:Kittel}. In section \ref{sec:FMRforNC}, we numerically solve the GP equations, obtaining resonance frequencies that depend upon the shape of the condensates, and spin currents driven by spin diffusion which is given by the MDDI.  Finally, Sec. \ref{sec:conclusion} is devoted to our conclusions.

\section{FORMULATION}\label{sec:formulation}

 In this section, we derive the equations of motion for spins from the spin-1 GP equations with an external magnetic field and an MDDI \cite{Yasunaga2008}.
\begin{eqnarray}\label{eq:GPoriginal}
i\hbar\frac{\partial \psi_\alpha}{\partial t} &=& \left(-\frac{\hbar^2}{2M}\nabla^2+V-\mu+c_0n \right)\psi_\alpha\nonumber \\
&&-g\mu_BH_iF^i_{\alpha \beta}\psi_\beta+c_2F_iF_{\alpha \beta}^i\psi_\beta \nonumber \\
&&+c_{{\rm dd}}\int d{\bf r}' \frac{\delta_{ij}-3e^ie^j}{|{\bf r}-{\bf r}'|^3}F_i({\bf r}')F_{\alpha \beta}^j\psi_\beta.
\end{eqnarray}
Here, $V$ is the trapping potential, $\mu$ is the chemical potential, and the total density $n = \sum _i n_i$ is given by $n _i = |\psi _i|^2$. The external magnetic field is ${\bf H} = (H_x, H_y, H_z)$, and the components $F_{\alpha \beta}^i$ of the spin matrices $\hat{F}_i$ are for spin-1. The interaction parameters are $c_0 = (g_0+2g_2)/3$ and $c_2 = (g_2-g_0)/3$ for $g_i = 4\pi\hbar^2a_i/M$ represented by s-wave scattering lengths $a_i$. The dipolar coefficient is $c_{{\rm dd}} = \mu_0g_e^2\mu_B^2/4\pi$, and the unit vector is ${\bf e} = (e^x, e^y, e^z) = (x-x', y-y', z-z')/|{\bf r}-{\bf r}'|$.

Under the homogeneous magnetic field ${\bf H} = H\hat{{\bf z}}$, the equations can be rewritten as,
\begin{subequations}\label{eq:GP}
\begin{eqnarray}
i\hbar\frac{\partial \psi _1}{\partial t} &=& \left(-\frac{\hbar ^2\nabla^2}{2M}+V-\mu +c_0n\right)\psi _1 -g\mu _BH\psi _1\nonumber\\
&+&c_2\{(n_1+n_0-n_{-1})\psi _1+\psi _{-1}^*\psi _0^2\}+D_1, \nonumber\\
\\
i\hbar\frac{\partial \psi _0}{\partial t} &=&\left(-\frac{\hbar ^2\nabla^2}{2M}+V-\mu +c_0n\right)\psi _0  \nonumber \\
&+&c_2\{(n_1+n_{-1})\psi _0+2\psi _0^*\psi _1\psi _{-1}\}+D_0, \nonumber\\
\\
i\hbar\frac{\partial \psi _{-1}}{\partial t} &=& \left(-\frac{\hbar ^2\nabla^2}{2M}+V-\mu +c_0n\right)\psi _{-1} +g\mu _BH\psi _{-1}\nonumber\\
&+&c_2\{(n_{-1}+n_0-n_1)\psi _{-1}+\psi _{1}^*\psi _0^2\}+D_{-1}. \nonumber\\
\end{eqnarray}
\end{subequations}
These dipolar terms are represented as,
\begin{eqnarray}
D_1 &=& \left(\frac{\psi _0}{\sqrt{2}}d_-+\psi _1d_z\right), \nonumber\\
D_0 &=& \left(\frac{\psi _1}{\sqrt{2}}d_++\frac{\psi _{-1}}{\sqrt{2}}d_-\right), \nonumber\\
D_{-1} &=& \left(\frac{\psi _0}{\sqrt{2}}d_+-\psi _{-1}d_z\right),\nonumber
\end{eqnarray}
with the integrations $d_\pm = d_x \pm id_y$ and $d_z$ given by, 
\begin{equation}\label{eq:int}
d _i = c_{{\rm dd}}\int d{\bf r}'\frac{F_i ({\bf r}')}{|{\bf r} -{\bf r}'|^3}\{1-3e^i\sum_{j}e^j\}.
\end{equation}
The spin density vectors $F_i$ are defined as,
\begin{subequations}\label{eq:f}
\begin{eqnarray}
F_x &=& {\bf \Psi}^\dagger\hat{F_x}{\bf \Psi} \nonumber \\
&=& \frac{\hbar}{\sqrt{2}}\{\psi _0^*(\psi _1+\psi _{-1})+\psi _0(\psi _1^*+\psi _{-1}^*)\}, \\
F_y &=& {\bf \Psi}^\dagger\hat{F_y}{\bf \Psi} \nonumber \\
&=&\frac{i\hbar}{\sqrt{2}}\{\psi _0^*(\psi _1-\psi _{-1})-\psi _0(\psi _1^*-\psi _{-1}^*)\},\\
F_z &=& {\bf \Psi}^\dagger\hat{F_z}{\bf \Psi} = \hbar(|\psi _1|^2-|\psi _{-1}|^2).
\end{eqnarray}
\end{subequations}
Here, ${\bf \Psi} = (\psi _1, \psi _0, \psi _{-1})^T$ is the spinor wave function. 

Differentiating Eq. \eqref{eq:f} with respect to time and utilizing Eq. \eqref{eq:GP}, we can obtain the Kittel-like equation,
\begin{equation}\label{eq:KGP}
\frac{\partial {\bf F}}{\partial t} = {\bf K} + \gamma _e[{\bf F} \times {\bf H}_{{\rm eff}}]
\end{equation}
with the gyromagnetic ratio $\gamma _e = g\mu _B/\hbar$ of an electron. The first term ${\bf K} = (K_x, K_y, K_z)$ becomes, 
\begin{eqnarray}
K_x &=& \frac{\hbar}{2Mi}\frac{1}{\sqrt{2}}\{(\psi _1 + \psi _{-1})\nabla^2\psi _0^*-\psi _0^*\nabla^2(\psi _1+\psi _{-1}) \nonumber\\
&&+\psi _0\nabla^2(\psi _1^*+\psi _{-1}^*)-(\psi _1^* + \psi _{-1}^*)\nabla^2\psi _0\}, 
\nonumber\\
K_y &=& \frac{\hbar}{2Mi}\frac{i}{\sqrt{2}}\{(\psi _1 - \psi _{-1})\nabla^2\psi _0^*-\psi _0^*\nabla^2(\psi _1-\psi _{-1}) \nonumber\\
&&-\psi _0\nabla^2(\psi _1^*-\psi _{-1}^*)+(\psi _1^* - \psi _{-1}^*)\nabla^2\psi _0\}, 
\nonumber\\
K_z &=& \frac{\hbar}{2Mi}(\psi _1\nabla^2\psi _1^*-\psi _1^*\nabla^2\psi _1 \nonumber \\
&&+\psi _{-1}^*\nabla^2\psi _{-1}-\psi _{-1}\nabla^2\psi _{-1}^*).\nonumber
\end{eqnarray}
The effective magnetic fields ${\bf H}_{{\rm eff}} = {\bf H} + {\bf H}_{{\rm dd}} = (H_{{\rm eff}}^x, H_{{\rm eff}}^y, H_{{\rm eff}}^z)$ consist of the external magnetic field  and the dipolar field ${\bf H}_{{\rm dd}}$, given by,
\begin{eqnarray}
H_{{\rm eff}}^x &=& -\frac{c_{{\rm dd}}}{g\mu _B}d_x, \nonumber\\
H_{{\rm eff}}^y &=& -\frac{c_{{\rm dd}}}{g\mu _B}d_y, \nonumber\\
H_{{\rm eff}}^z &=& H-\frac{c_{{\rm dd}}}{g\mu _B}d_z. \nonumber
\end{eqnarray}
Note that Eq. \eqref{eq:KGP} does not depend on spin exchange interaction, which refers to  the second term with $c_2$ in Eq. \eqref{eq:GPoriginal}. Generally, the interaction affects a spin through the effective magnetic fields of the other spins. However, exchange interaction does not appear in ${\bf H}_{{\rm eff}}$. Therefore, the isotropic exchange interaction does not affect MR in these condensates. 

We can redefine Eq. \eqref{eq:KGP} as, 
\begin{equation}\label{eq:KGP2}
\frac{\partial F_k}{\partial t} = \frac{\hbar}{2Mi}\nabla^2 F_k -\nabla\cdot{\bf j}_k +\gamma _e [{\bf F} \times {\bf H}_{{\rm eff}}]_k,
\end{equation}
where, 
\begin{eqnarray}
{\bf j}_x &=& \frac{\hbar}{\sqrt{2}Mi}(\psi _0^*\nabla(\psi _1+\psi _{-1})+(\psi _1^*+\psi _{-1}^*)\nabla\psi _0), \nonumber \\
{\bf j}_y &=&\frac{\hbar}{\sqrt{2}M}(\psi _0^*\nabla(\psi _1-\psi _{-1})-(\psi _1^*-\psi _{-1}^*)\nabla\psi _0), \nonumber \\
{\bf j}_z &=& \frac{\hbar}{Mi}(\psi _1^*\nabla\psi _1 -\psi _{-1}^*\nabla\psi _{-1}). \nonumber 
\end{eqnarray}
The equation of motion \eqref{eq:KGP2} for spins describes the properties of spin dynamics in a ferromagnetic fluid. The first, second, and third terms of Eq. \eqref{eq:KGP2} represent spin diffusion, spin current, and spin precession around ${\bf H}_{{\rm eff}}$, respectively. 

 Comparing Eq. \eqref{eq:KGP2} with Eq. \eqref{eq:Kittel}, we noticed several differences. First, Eq. \eqref{eq:KGP2} was directly derived from the GP equations, whereas Eq. \eqref{eq:Kittel} is a phenomenological  equation of magnetization. The spin density vectors in Eq. \eqref{eq:KGP2} are microscopically affected by other spins through the dipolar fields in the effective magnetic fields. On the other hand, the magnetization in Eq. \eqref{eq:Kittel} is affected by demagnetizing fields originating from macroscopically polarized magnetization in the condensed matter. Namely, Eq. \eqref{eq:KGP2} can describe the macroscopic demagnetizing field resulting from the microscopic dipolar field. This is a very important difference between these equations. 

We initially investigated the physics of the first and second terms of Eq. \eqref{eq:KGP2}. To simplify the discussion, we considered the equation under the condition ${\bf H}_{{\rm eff}} ={\bf 0}$. Thus, we derived the continuity equations,
\begin{equation}
\frac{\partial F_i}{\partial t} +\nabla \cdot {\bf J}_i = 0,\label{eq:con}\\
\end{equation}
where ${\bf J}_k = {\bf j}_k-\hbar/(2Mi)\nabla F_k$ is an effective current term,
\begin{subequations}\label{eq:spincurrents}
\begin{eqnarray}
{\bf J}_x &=& -\frac{i\hbar^2}{2\sqrt{2}M}\{\psi _0^*\nabla(\psi _1+\psi _{-1})+(\psi _1^*+\psi _{-1}^*)\nabla\psi _0 \nonumber \\&&-\psi _0\nabla(\psi _1^*+\psi _{-1}^*)-(\psi _1+\psi _{-1})\nabla\psi _0^*\},\\
{\bf J}_y &=& \frac{\hbar^2}{2\sqrt{2}M}\{\psi _0^*\nabla(\psi _1-\psi _{-1})-(\psi _1^*-\psi _{-1}^*)\nabla\psi _0 \nonumber \\&&+\psi _0\nabla(\psi _1^*-\psi _{-1}^*)-(\psi _1-\psi _{-1})\nabla\psi _0^*\},\\
{\bf J}_z &=& -\frac{i\hbar^2}{2M}(\psi _1^*\nabla\psi _1-\psi _1\nabla\psi _1^* -\psi _{-1}^*\nabla\psi _{-1}+\psi _{-1}\nabla\psi _{-1}^*).\nonumber\\
\end{eqnarray}
\end{subequations}
 Equation \eqref{eq:con} can also be rewritten as,
\begin{equation}
\frac{d}{dt}\int _VF_idV = \int _V \nabla\cdot{\bf J}_idV = \int _S {\bf J}_i\cdot {\bf n}dS,\nonumber
\end{equation} 
by using the volume integral and the surface integral, whose unit vector ${\bf n}$ is vertical to the surface for  Stokes' theorem. The equation indicates that the expectation value of the spin matrix $\langle \hat{F}_i \rangle = \int dV F_i$ in the volume $V$ is conserved for the spin probability flux ${\bf J}_i$ leaving and entering the surface.

Under ${\bf H}_{{\rm eff}} \neq 0$, the Kittel-like equation can be reduced to the following equation,
\begin{equation}\label{eq:conmddi}
\frac{\partial F_i}{\partial t} +\nabla \cdot {\bf J}_i = [ {\bf F}\times {\bf H}_{{\rm eff}}]_i,
\end{equation}
 where the right side of the equation breaks the conservation law of spin density. Therefore, the Kittel-like equations have two dynamics: spin precessions with frequency given by the effective magnetic field and spin currents without spin conservation. The spin currents of the system will be discussed in Sec. \ref{subsec:spincurrent}
 
\section{FMR under single-mode approximation}\label{sec:FMRunderSMA} 

In order to study the basic properties of the second term in Eq. \eqref{eq:KGP}, we introduced the single-mode approximation, 
\begin{equation}\label{eq:SMA}
\psi_i({\bf r},t) = \sqrt{N}\xi_i(t)\phi({\bf r})\exp\left(-\frac{i\mu t}{\hbar}\right),
\end{equation}
where $\phi$ satisfies the eigenvalue equation $(-\hbar^2\nabla ^2/2M+V+c_0n)\phi = \mu \phi$ with the relation $\int d{\bf r} |\phi|^2 = 1$. The approximation is effective when the shapes of the condensates are determined by the spin-independent terms, namely $|c_0| \gg |c_2|$ \cite{Zhang2005}. For $^{87}$Rb and $^{23}$Na, the relation is satisfied. Under this approximation, the first term of Eq. \eqref{eq:KGP} vanishes, and we obtain the Kittel equation for the spatially independent spin density vector ${\bf S} = (S_x, S_y, S_z)$,
\begin{equation}\label{eq:K}
\frac{d {\bf S}}{dt} = \gamma _e[{\bf S} \times {\bf H}_{{\rm eff}}^{{\rm SMA}}],
\end{equation}
  where,
\begin{eqnarray}
S_x &=& \frac{\hbar}{\sqrt{2}}\{\xi _0^*(\xi _1 +\xi _{-1})+\xi _0(\xi _1^*+\xi _{-1}^*)\}, \nonumber\\
S_y &=& \frac{i \hbar}{\sqrt{2}}\{\xi _0^*(\xi _1 -\xi _{-1})-\xi _0(\xi _1^*-\xi _{-1}^*)\}, \nonumber\\
S_z &=& \hbar(|\xi _1|^2-|\xi _{-1}|^2),\nonumber
\end{eqnarray}
and the effective magnetic field ${\bf H}_{{\rm eff}}^{{\rm SMA}} = (-N_{{\rm dd}}^xS_x, -N_{{\rm dd}}^yS_y, H-N_{{\rm dd}}^zS_z)$ is given by
\begin{equation}\label{eq:cdd}
N_{{\rm dd}}^i = \frac{c_{{\rm dd}}}{g\mu _B}N\int\int d{\bf r}d{\bf r}'\frac{|\phi({\bf r})|^2|\phi({\bf r}')|^2}{|{\bf r}-{\bf r}'|^3}\{1-3e^i\sum _je^j\}.
\end{equation}
Equation \eqref{eq:K} also indicates that the spin vector ${\bf S}$ precesses around ${\bf H}_{{\rm eff}}^{{\rm SMA}}$. The precession frequency reveals the characteristic dynamics.  Next, we consider a small deviation $\delta {\bf S} = (\delta S_x, \delta S_y, \delta S_z)$ around the stationary solution, ${\bf S}_0 = S_0\hat{{\bf z}}$ with $S_0 = H_0/N_{{\rm dd}}^z$, of Eq. \eqref{eq:K}, namely ${\bf S} = {\bf S}_0+\delta {\bf S}$. Introducing this representation into Eq. \eqref{eq:K} and linearizing the equation, we derived the following equations,
\begin{eqnarray}
\frac{d}{dt}\delta S_x &=& \gamma _e\{H+(N_{{\rm dd}}^y-N_{{\rm dd}}^z)S_0\}\delta S_y, \nonumber\\
\frac{d}{dt}\delta S_y &=& -\gamma _e\{H+(N_{{\rm dd}}^x-N_{{\rm dd}}^z)S_0\}\delta S_x, \nonumber\\
\frac{d}{dt}\delta S_z &=& 0,\nonumber
\end{eqnarray} 
which give the resonance frequency,
\begin{equation}\label{eq:wdd}
\omega^2 = \gamma _e^2 \{H+(N_{{\rm dd}}^x-N_{{\rm dd}}^z)S_0\}\{H+(N_{{\rm dd}}^y-N_{{\rm dd}}^z)S_0\}
\end{equation}
The spin precesses with the resonance frequency $\omega$, which depends on the dipolar terms $N_{{\rm dd}}^i$. 

Here, we consider the single particle density distribution $|\phi ({\bf r})|^2 \propto e^{-(x^2+y^2+\lambda _zz^2)/a^2}$, where $\lambda _z$ is the aspect ratio, and discuss simple situations. For the spherical case of $\lambda _z = 1$, the integration \eqref{eq:cdd} results in $N_{{\rm dd}}^x = N_{{\rm dd}}^y = N_{{\rm dd}}^z$, giving $\omega = \gamma _e H$. The dipolar fields are canceled because of the isotropy, so that the spin precesses with Larmor frequency. For the circular plane (infinite cylinder) case of $\lambda _z = \infty \ (0)$, we obtain $\omega = \gamma _e\{H-(N_{{\rm dd}}^x-N_{{\rm dd}}^z)S_0\}$ for $N_{{\rm dd}}^x = N_{{\rm dd}}^y$.  

In this representation, it seems that the microscopic dipolar fields, Eq. \eqref{eq:cdd}, act as a macroscopic demagnetizing field to compare  Eq. \eqref{eq:wk} with \eqref{eq:wdd}. We believe that the origin of the demagnetizing field is an MDDI. If the above discussion is correct, the dipolar coefficients $N_{{\rm dd}}^i$ should depend on the shape of the condensates. However, the single-mode approximation in spinor dipolar BECs is not effective in large-aspect-ratio condensates, as discussed by Yi and Pu \cite{Yi2006}. Therefore, we must consider the spin dynamics beyond the approximation. 

\section{FMR for Numerical Calculation}\label{sec:FMRforNC} 

\subsection{Precession dependence on the aspect ratio $\lambda$}\label{subsec:precession}

In this section, we discuss FMR by numerically calculating the two-dimensional Eq. \eqref{eq:GPoriginal} under the condition of $^{87}$Rb, namely $c_0 \gg -c_2 > 0$. We began calculating the spin precessions by applying a $\pi/20$ pulse to the ground state, whose spins were polarized to the uniform magnetic field ${\bf H} = H\hat{{\bf z}}$ trapped by $V = M\omega _x^2(x^2+\lambda^2y^2)/2$ with $g\mu _B H/\hbar \omega _x = 20$ and an aspect ratio $\lambda = \omega _y/\omega _x$.

We investigated the dynamics of $\langle F_x \rangle$ for $\lambda = 0.5, 1$, and $1.5$ with and without the MDDI. From $t = 0$ to $\pi/(20\gamma _eH)$, a $\pi/20$ pulse was applied. Then, the spins were tilted by $\pi/20$ radians from the $z$ axis with precession. After turning off the pulse, the spins precessed around the $z$ axis, conserving $\langle F_z \rangle$.  We define the notation $\langle F_i \rangle _{\lambda = \lambda _a} ^{dd}$ and $\langle F_i \rangle _{\lambda = \lambda _a}$ as indicating the expectation values of $F_i$ with and without an MDDI in the trap with $\lambda = \lambda _a$.

\begin{figure}[t]
\begin{center}
\includegraphics[width=0.9\linewidth]{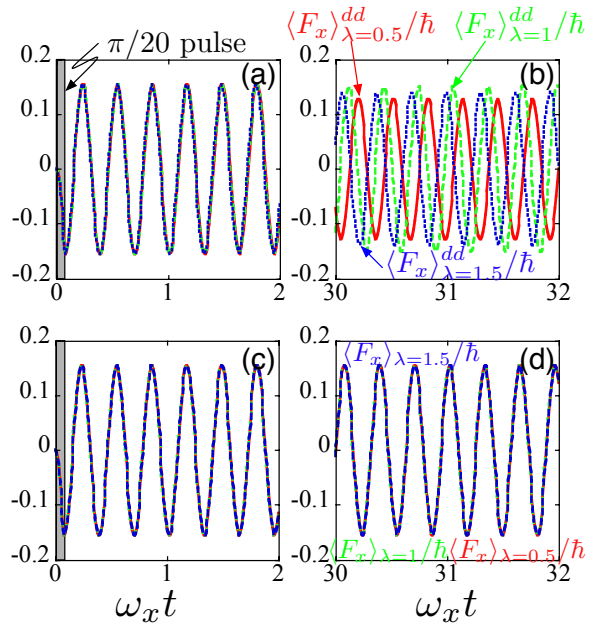} 
\caption{(Color online) The time development of $\langle F_x\rangle _{\lambda}^{dd}$, (a) and (b), and $\langle F_x\rangle _{\lambda}$, (c) and (d). The red solid, blue dashed, and green dotted lines show the results of $\lambda = 0.5$, $1$, and $1.5$ respectively. The gray zone represents the duration of a $\pi/20$ pulse.}
\label{fig:lamdepend}
\end{center}
\end{figure}

First, the typical motions of spins are shown in Fig. \ref{fig:lamdepend}. Investigating the time development of $\langle F_i \rangle _{\lambda = 0.5}^{dd}$, $\langle F_i \rangle _{\lambda = 1}^{dd}$, and $\langle F_i \rangle _{\lambda = 1.5}^{dd}$, we obtained the differences between their precession frequencies, as shown in Fig. \ref{fig:lamdepend} (a) and (b). The differences appeared at frequencies below the Larmor frequency, given by $H$. For $0 \leq t \geq 2 $,  no deviation between the precessions was observed, but deviations clearly appeared as more time elapsed. In order to demonstrate that the $\lambda$ dependence was given not by $H$ but by $H_{{\rm dd}}$, we show precessions for the same aspect ratios without the MDDI in Fig. \ref{fig:lamdepend} (c) and (d). The precession frequency did not change without the MDDI for different values of $\lambda$. Therefore, the dipolar frequency $\omega _{{\rm dd}} =\gamma _eH_{{\rm dd}}$ depends upon the shape of the condensate. 
%LaTeX Warning: A float is stuck (cannot be placed) on input line 277
\begin{figure}[t]
\begin{center}
\includegraphics[width=0.9\linewidth]{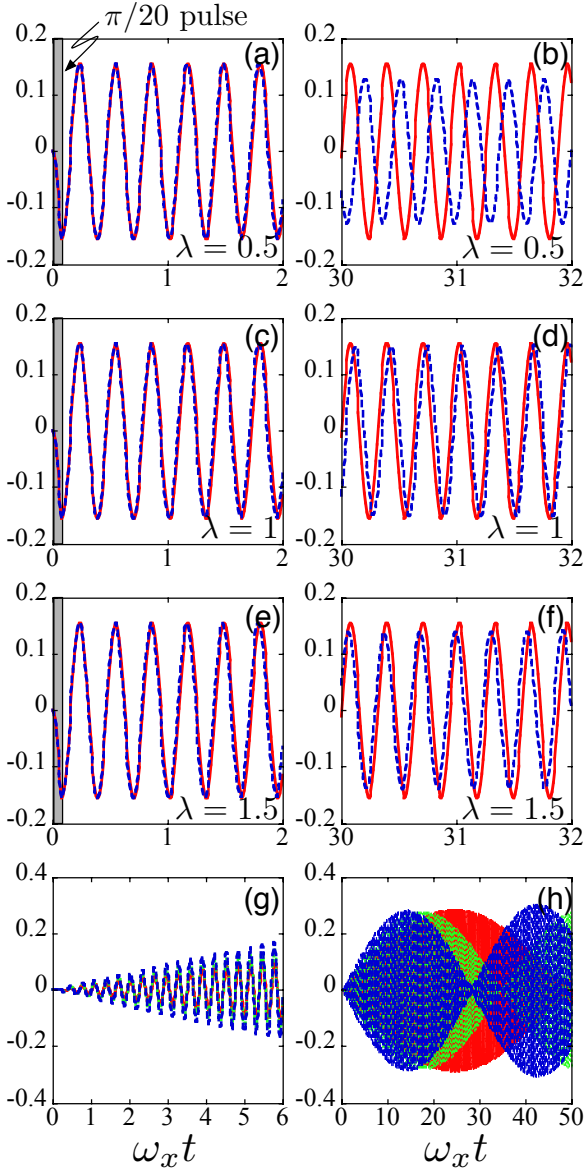} 
\caption{(Color online) Comparing the precession with and without the MDDI. (a) and (b), (c) and (d), and (e) and (f) show the precession for $\lambda = 0.5$, $1$, and $1.5$, respectively. The solid and dashed lines are $\langle F_x\rangle _\lambda$ and $\langle F_x\rangle _\lambda^{dd}$. (g) and (h) represent $(\langle F_x\rangle _{\lambda = 0.5}^{dd}-\langle F_x\rangle _{\lambda = 0.5})/\hbar$ (solid), $(\langle F_x\rangle _{\lambda = 1}^{dd}-\langle F_x\rangle _{\lambda = 1})/\hbar$ (dot), and $(\langle F_x\rangle _{\lambda = 1.5}^{dd}-\langle F_x\rangle _{\lambda = 1.5})/\hbar$ (dashed), respectively.}
\label{fig:cddepend}
\end{center}
\end{figure}

\begin{figure}[t]
\begin{center}
\includegraphics[width=0.9\linewidth]{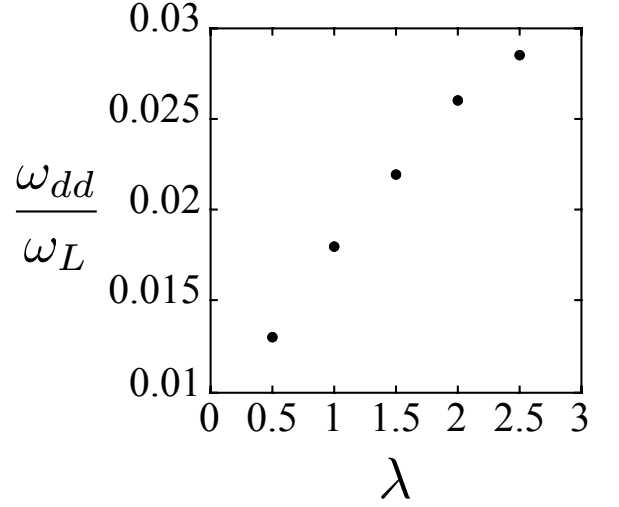} 
\caption{$\lambda$ dependence of $\omega _{{\rm dd}}/\omega _L$.}
\label{fig:omega_lambda}
\end{center}
\end{figure}
	
Next, we examined the effects of the MDDI on the precessions in Fig. \ref{fig:cddepend}. Comparing $\langle F_x \rangle _{\lambda}^{dd}$ with $\langle F_x \rangle _{\lambda}$, we observed that the MDDI caused an effective magnetic field, because the frequency of the precession with the MDDI deviated from that without the MDDI in Fig. \ref{fig:cddepend} (a) to (f). Assuming that $\langle F_i \rangle _{\lambda = 1}^{dd}-\langle F_i \rangle _{\lambda = 1}$ is represented approximately to $A\cos\gamma _e(H+H_{{\rm dd}})t-A\cos\gamma _eHt$ with an amplitude $A$, we extracted the dipole frequency from the waveform. Since the waveform became $-2A\sin \omega _{{\rm dd}}t/2\sin(\omega _L+\omega _{{\rm dd}}/2)t$, the beat consisted of the large frequency $\omega _L + \omega _{{\rm dd}}/2$ and the small frequency $\omega _{{\rm dd}}/2$. From Fig. \ref{fig:cddepend} (h), we estimated these frequencies to obtain $\omega _{{\rm dd}}/\omega _{L} \simeq 6.5, 9, and 11 \times 10^{-3}$ for $\lambda = 0.5, 1, and 1.5$ respectively.

Figure \ref{fig:omega_lambda} shows the $\lambda$ dependence of $\omega _{{\rm dd}}/\omega _L$. From the results, however, we cannot safely conclude that the $\lambda$ dependence of the frequencies is given by changing the shape  of the condensates, since the dipolar frequencies may be given by change of the density with  the shape. FMR in condensed matters has been discussed in condensed matter of uniform density, even with changing shape. On the other hand, atomic BECs have tunable density and shape. Therefore, our calculations indicate characteristic of  FMR in atomic cold gases.    

\begin{figure}[t]
\begin{center}
\includegraphics[width=0.9\linewidth]{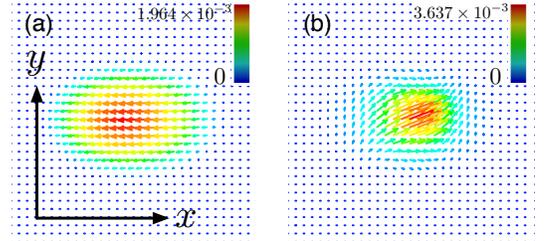} 
\caption{(Color online) Projection of ${\bf F}$ onto the $x$ - $y$ plane for $\lambda = 1.5$ and $\omega _xt = 12.7$. The figures show the results (a) with MDDI and (b) without that. The vectors are nondimensionalized.  }
\label{fig:rdepend}
\end{center}
\end{figure}

\subsection{Spin current}\label{subsec:spincurrent}

We observed spin currents driven by spin diffusion, which was caused by a ${\bf r}$ dependence of the dipolar field. Figure \ref{fig:rdepend} shows the projections of ${\bf F}$ onto the $x$ - $y$ plane for $\lambda = 1.5$ and $\omega _xt = 12.7$. The precession with the MDDI lost homogeneity of the spin directions, whereas the precession without the MDDI maintained this homogeneity. This is because the precession frequency has an {\bf r} dependence, specifically, $\omega ({\bf r}) = \gamma _e H_{{\rm eff}}({\bf r}) = \gamma _e (H+H_{{\rm dd}}({\bf r}))$. 

\begin{figure}[t]
\begin{center}
\includegraphics[width=0.9\linewidth]{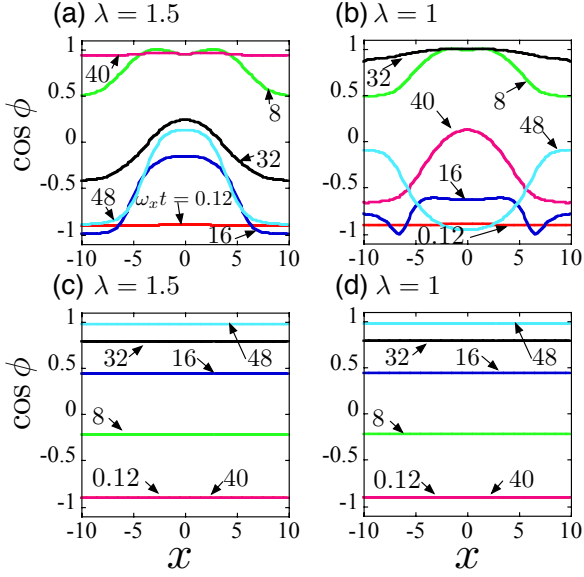} 
\caption{Dynamics of a cross-section of $F_x/|F_{xy}|$ at $y=0$, where $|F_{xy}| = \sqrt{F_x^2+F_y^2}$. From the relation $F_x = |F_{xy}|\cos\phi$, the parameter represents $\cos\phi$. The results with the MDDI (a) and without it (b) are shown for $\lambda = 1.5$, and (c) and (d) show results for $\lambda = 1$.  The $x$ axis are nondimensionalized by $\sqrt{\hbar/M\omega _x}$}
\label{fig:cs_angle}
\end{center}
\end{figure}

The dipole interaction drives the spin diffusion, which is shown in Fig. \ref{fig:cs_angle}. The figure shows $F_x/|F_{xy}| = \cos\phi$ as a function of $x$ at $y = 0$, where $\phi$ is the angle between the spin vector and the $x$ axis. In the dynamics with the dipole interaction for $\lambda = 1.5$ (a) and $1$ (b), the spin densities lost their angular coherence,  whereas the dynamics without the dipole interactions maintained this coherence ( (c) and (d)).

\begin{figure}[t]
\begin{center}
\includegraphics[width=0.9\linewidth]{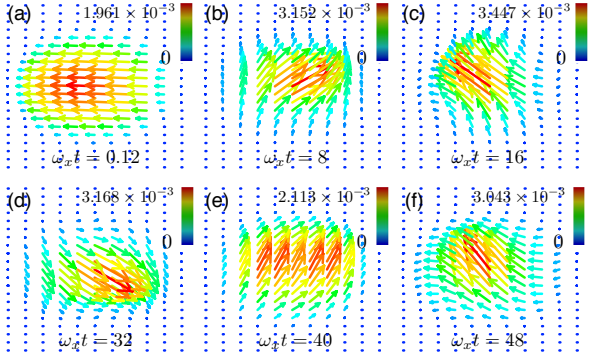} 
\caption{(Color online) Dynamics of ${\bf F}$ projected onto the $x$ - $y$ plane for $\lambda = 1.5$ with dipolar interaction.}
\label{fig:dynamics_cd02}
\end{center}
\end{figure}

The spin diffusion drives the spin current ${\bf J}_k$ in Eq. \eqref{eq:spincurrents}, which is shown in Fig. \ref{fig:dynamics_cd02}. In order to explain how the spin current is driven by the spin diffusion, we considered the amplitudes of the wave functions $\psi _j = f_j e^{i\varphi _j}$ as, 
\begin{subequations}\label{eq:wavefunction}
\begin{eqnarray}
\psi _1 ({\bf r},t) &=& \frac{\sqrt{n ({\bf r},t)}}{2}(1+\cos\theta({\bf r},t))e^{i\varphi _1 ({\bf r},t)}, \nonumber\\ \\
\psi _0 ({\bf r},t) &=& \sqrt{\frac{n ({\bf r},t)}{2}}\sin\theta({\bf r},t)e^{i\varphi _0 ({\bf r},t)}, \\
\psi _{-1} ({\bf r},t) &=& \frac{\sqrt{n ({\bf r},t)}}{2}(1-\cos\theta({\bf r},t))e^{i\varphi _{-1} ({\bf r},t)}, \nonumber \\
\end{eqnarray}
\end{subequations}
where the forms show the ground state of the ferromagnetic state \cite{Ho1998}. The amplitude is represented by $n$ and the angle $\theta$ between the spin and the $z$ axis. We introduced this representation to demonstrate that the spin current is derived from the spin diffusion. Of course, we confirmed the validity of the ferromagnetic representation under the pulse and magnetic field by calculating $\theta$ directly. Therefore, it can be utilized for the polarized spin state studied in our work. The amplitudes $f_{\pm 1}$ were formed to represent $F_z = n\hbar \cos\theta$, and $f_0$ was determined to satisfy the relation $n = \sum_j |\psi _j|^2$. For example, $(n_1, n_0, n_{-1}) = (n, 0, 0)$ led to $F_z = n\hbar$ with $\theta = 0$, and $(n_1, n_0, n_{-1}) = (n/4, n/2, n/4)$ resulted in $F_z = 0$ with $\theta = \pi/2$. The wave function can only express the ferromagnetic states, {\it i.e.} the form  cannot represent the antiferromagnetic state $(n_1, n_0, n_{-1}) = (n/2, 0, n/2)$ or the polar state $(n_1, n_0, n_{-1}) = (0, n, 0)$. This restriction of the wave function is caused by the first representation $F_z = n\hbar \cos\theta$. 

\begin{figure}[t]
\begin{center}
\includegraphics[width=0.9\linewidth]{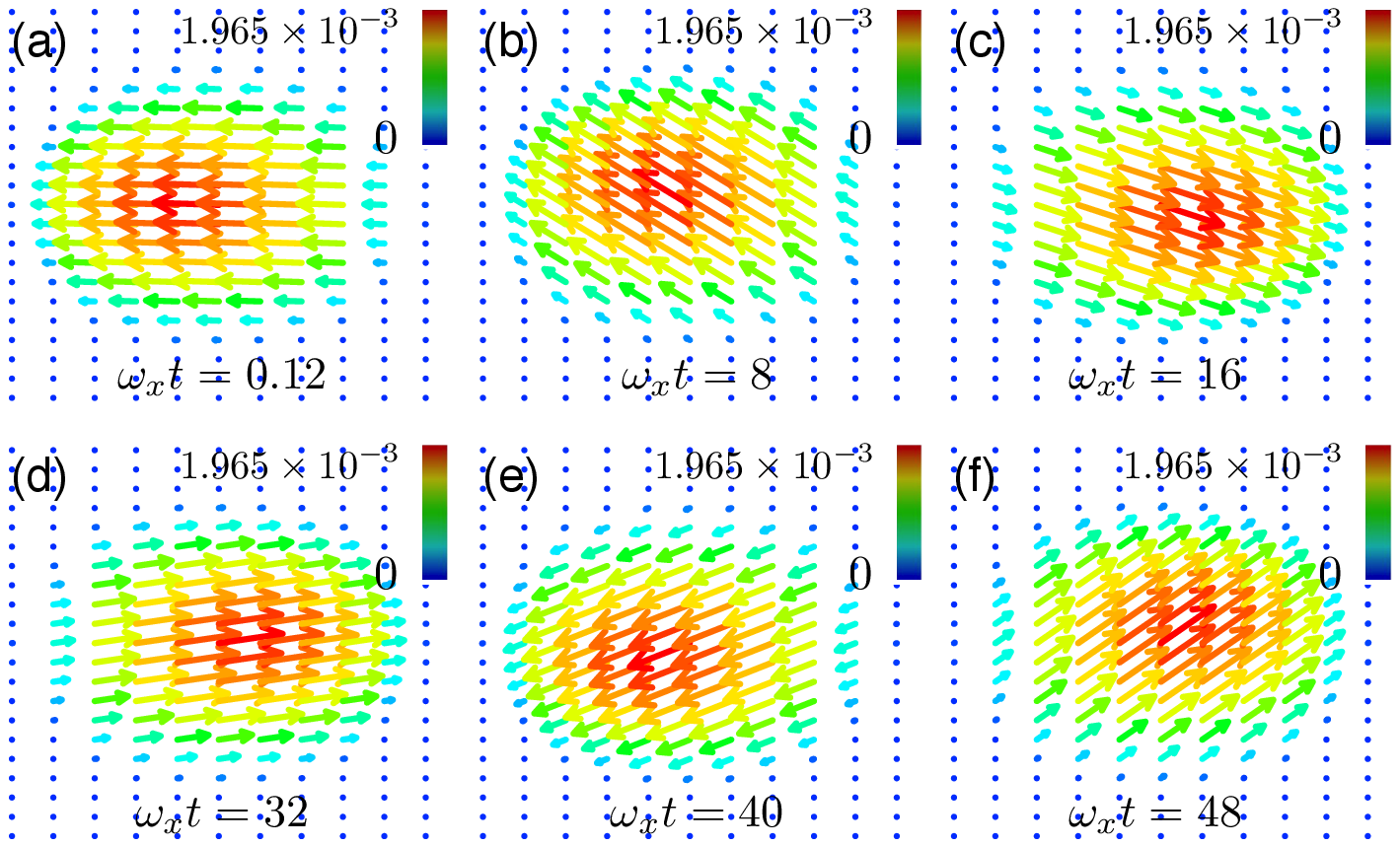} 
\caption{(Color online) Dynamics of ${\bf F}$ projected onto the $x$ - $y$ plane for $\lambda = 1.5$ without dipolar interaction.}
\label{fig:dynamics_cd0}
\end{center}
\end{figure}

By introducing this representation into Eqs. \eqref{eq:f} and \eqref{eq:spincurrents}, we can redefine as follows,
\begin{eqnarray}
F_x &=& n\hbar \sin\theta\left(\cos\varphi _{\rm r}\cos\varphi-\cos\theta\sin\varphi _{\rm r}\sin\varphi\right), \nonumber\\
F_y &=& -n\hbar \sin\theta\left(\cos\varphi _{\rm r}\sin\varphi+\cos\theta\sin\varphi _{\rm r}\cos\varphi\right), \nonumber
\end{eqnarray} 
and, 
\begin{subequations}\label{eq:spincurrents2}
\begin{eqnarray}
{\bf J}_x &=& \frac{n\hbar^2}{4M}\bigg\{\sin\theta(1+\cos\theta)\cos(\varphi _1-\varphi _0)\nabla\varphi _1\nonumber\\&+&\sin\theta(1-\cos\theta)\cos(\varphi _{-1}-\varphi _0)\nabla\varphi _{-1}\nonumber \\&-&2\sin\theta\left(\cos\varphi _{\rm r}\cos\varphi-\cos\theta\sin\varphi _{\rm r}\sin\varphi\right)\nabla\varphi _0\nonumber\\&+&2\left(\cos\varphi _{\rm r}\sin\varphi+\cos\theta\sin\varphi _{\rm r}\cos\varphi\right)\nabla\theta\bigg\},\\
{\bf J}_y &=&-\frac{n\hbar^2}{4M}\bigg\{\sin\theta(1+\cos\theta)\sin(\varphi _1-\varphi _0)\nabla\varphi _1\nonumber\\&-&\sin\theta(1-\cos\theta)\sin(\varphi _{-1}-\varphi _0)\nabla\varphi _{-1}\nonumber\\&+&2\sin\theta\left(\cos\varphi _{\rm r}\sin\varphi+\cos\theta\sin\varphi _{\rm r}\cos\varphi\right)\nabla\varphi _0\nonumber\\&+&2\left(\cos\varphi _{\rm r}\cos\varphi-\cos\theta\sin\varphi _{\rm r}\sin\varphi\right)\nabla\theta\bigg\},\\
{\bf J}_z &=&\frac{n\hbar^2}{4M}\{(1+\cos\theta)^2\nabla\varphi _1-(1-\cos\theta)^2\nabla\varphi _{-1}\},\nonumber \\
\end{eqnarray}
\end{subequations}
where $\varphi _{\rm r} = (\varphi _1 + \varphi _{-1}-2\varphi _0)/2$ and $\varphi = (\varphi _1-\varphi _{-1})/2$ are relative phases. Since the relation $\varphi _{\rm r} = 0$ was satisfied in our calculations, we used the relation in Eqs. \eqref{eq:spincurrents2}, and the spin density vector formed an azimuthal angle $\varphi$ with the $x$ axis. Then, we derived the spin components $F_x = n\hbar\cos\varphi\sin\theta$, $F_y = n\hbar\cos\varphi\sin\theta$, and $F_z = n\hbar\cos\theta$. We can therefore rewrite the spin density currents,
\begin{subequations}
\begin{eqnarray}\label{eq:currents}
{\bf J}_x &=& \frac{n\hbar^2}{4M}(4\cos\varphi\sin\theta\nabla\varphi _0 \nonumber\\&&+ 2\cos\varphi\sin\theta\cos\theta\nabla\varphi - 2\sin\varphi\nabla\theta), \\
{\bf J}_y &=& -\frac{n\hbar^2}{4M}(4\sin\varphi\sin\theta\nabla\varphi _0 \nonumber\\&&+ 2\sin\varphi\sin\theta\cos\theta\nabla\varphi + 2\cos\varphi\nabla\theta),\\
{\bf J}_z &=& \frac{n\hbar^2}{4M}\{4\cos\theta\nabla\varphi _0 + 4(1+\cos^2\theta)\nabla\varphi\},
\end{eqnarray}
\end{subequations}
which are driven by the gradients of the angles, $\varphi$ and $\theta$, and the phase $\varphi _0$. In the  precessions with MDDI, the gradients occurred because of the dipolar fields ${\bf H}_{{\rm dd}} ({\bf r})$. As a result, the spin currents were clearly driven, as shown in Fig. \ref{fig:dynamics_cd02}. For $\omega _x t =0.12$, the spin vectors were coherent just after the applied $\pi/20$ pulse (Fig. \ref{fig:dynamics_cd02} (a)). The spin densities, $F_x$ and $F_y$, then flowed to the center of the condensates from Fig. \ref{fig:dynamics_cd02} (b) to (c). Then, the densities reversed, and diffused outward from Fig. \ref{fig:dynamics_cd02} (d) to (e). This oscillation was repeated. Of course, we cannot obtain the spin current without the dipolar interactions, since the gradients of $\theta $ and $\varphi$ were not caused; the dynamics are shown in Fig. \ref{fig:dynamics_cd0}.

\begin{figure}[t]
\begin{center}
\includegraphics[width=0.9\linewidth]{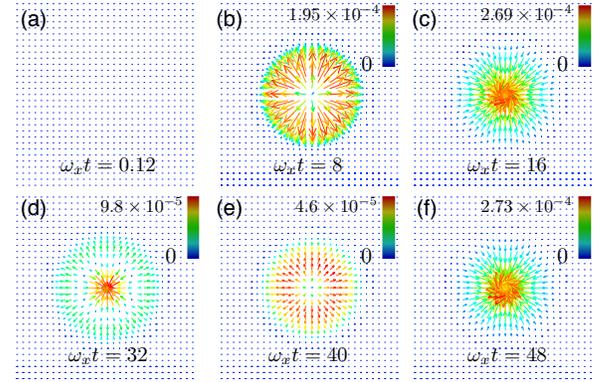} 
\caption{(Color online) Dynamics of the spin currents ${\bf J}_x$ projected onto the $x-y$ plane for $\lambda = 1$ with dipolar interaction. The vectors are nondimensionalized.}
\label{fig:scdl1}
\end{center}
\end{figure}

\begin{figure}[t]
\begin{center}
\includegraphics[width=0.9\linewidth]{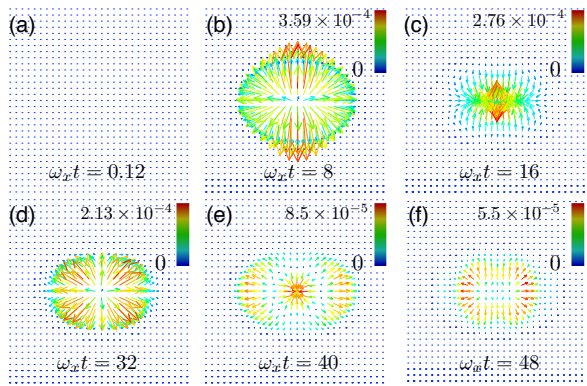} 
\caption{(Color online) Dynamics of the spin currents ${\bf J}_x$ projected onto the $x-y$ plane for $\lambda = 1.5$ with dipolar interaction.}
\label{fig:scdl15}
\end{center}
\end{figure}

\begin{figure}[thp]
\begin{center}
\includegraphics[width=0.9\linewidth]{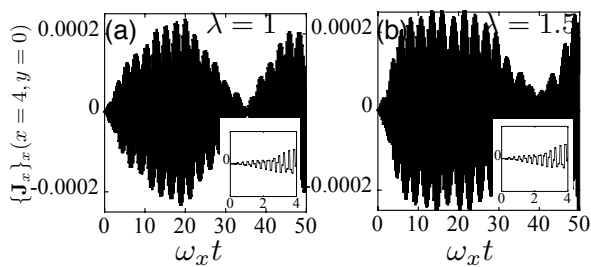} 
\caption{Dynamics of the $x$ component of ${\bf J}_x$ at $x=4$ and $y=0$. The inter figures are the results for $\omega _x t = 0$ to $4$.}
\label{fig:sctimedevelop}
\end{center}
\end{figure}

In order to investigate the spin fluid dynamics, we calculated the spin current ${\bf J}_x$ for Eq. \eqref{eq:spincurrents2}, as shown in Figs. \ref{fig:scdl1} and \ref{fig:scdl15}. These figures represent ${\bf J}_x$ from the previous calculations with $\lambda = 1$ and $1.5$ respectively. Despite the difference in the ratio, we observed two common properties in these figures. The direction of the currents changed rapidly, corresponding to the large precession frequency, and the magnitudes changed slowly with the small dipolar frequency, as shown in Fig. \ref{fig:sctimedevelop}, which shows the time development of the $x$ component of ${\bf J}_x (x=4, y=0)$.  This figure indicates that the oscillation of the current direction occurred with the precession frequency, which varied in magnitude with changing dipolar frequency. Eq. \eqref{eq:conmddi} also indicates that the spin density was not conserved because of the effective magnetic field. Therefore, the spin currents can be driven from a source and sink in the center of the condensates, as in Figs. \ref{fig:scdl1} and \ref{fig:scdl15}.  The two common properties were insensitive to the value of $\lambda$. However, the change in spin density for $\lambda = 1.5$ exhibited quadratic pole motion in a scissors-like mode for mass density \cite{PethickSmith}, which can be understood as an oscillation between the spin density migrating to the $y$ axis from the $x$ axis and back again, as shown in Figs. \ref{fig:dynamics_cd02} (a) to (c). Therefore, the spin collective mode was caused by spin diffusions induced  by the MDDI. Therefore, the spin current causes the dynamics of spin scissors-like mode, which was observed as a shrinking and expansion of the spin density in Fig. \ref{fig:dynamics_cd02}. The shrinking and expansion were common features for $\lambda = 1$ and $1.5$. However, the spin currents were affected by the symmetry of the traps, as shown in Figs. \ref{fig:scdl1} and \ref{fig:scdl15}.

From the calculations, we expected that the spin current would be observable when using the spinor BECs. Recently, spin current is focused from fields of spintronics. However, it is difficult to observe the spin current in metals and condensed matter. Atomic BECs, a macroscopic quantum phenomenon, can show the spin current clearly and directly in the dynamics of the spinor densities. Therefore, we should attempt to observe various spin currents utilizing tunable experimental parameters, {\it i.e.} interaction parameters, trap frequencies, and the number of particles.   

%LaTeX Warning: A float is stuck (cannot be placed) on input line 370.

%LaTeX Warning: A float is stuck (cannot be placed) on input line 371

%LaTeX Warning: A float is stuck (cannot be placed) on input line 372.

%LaTeX Warning: A float is stuck (cannot be placed) on input line 373

\section{CONCLUSION}\label{sec:conclusion}

We investigated the properties of magnetic resonance in spinor dipolar BECs by calculating the GP equations, obtaining Kittel-like equations as the equations of motion for the spin density vector. The equations revealed two properties. One is the dynamics of the spin fluid, and the other is precession under the effective magnetic field consisting of the external magnetic fields and the dipolar fields. The magnetic resonance with the properties of the spin fluid  was characteristic of this system. 

In order to extract properties from the GP equations, we studied the law of conservation of spin density current without effective magnetic fields by first deriving the continuity equations from the GP equation, obtaining representations of the spin current. Second, we analytically evaluated the precession dynamics described by the Kittel equations derived from the GP equations using a single-mode approximation, where the Kittel equations show conventional FMR.  The analysis clearly indicated that the origin of the FMR in the BECs is like  the dipolar field, whereas the origin of the resonance in the Kittel equations for condensed matter is the demagnetizing field. Comparing the FMR of the BEC with that of the condensed matter, we concluded that the origin of the resonance was not the spin exchange interaction that causes magnetism in condensed matter, but the anisotropy of the MDDI. Finally, we numerically calculated the GP equations, representing the dynamics with the two common properties.  The characteristic dynamics showed that the effective magnetic field introduced spin diffusion into the Larmor precession, driving the spin-current-like scissors modes. 

The relation between the spin current and FMR has not yet been discussed for typical FMR. Therefore, it is important to study spin current in condensates. We also believe that the study of spin current will be useful for the development of spintronics, because it is difficult to directly observe spin currents in condensed matter spintronics.

\section{ACKNOWLEDGMENT}\label{sec:ack}

M. Y. acknowledges the support of a Research Fellowship of the Japan Society for the Promotion of Science for Young Scientists (Grant No. 209928). M. T. acknowledges the support of a Grant-in Aid for Scientific Research from JSPS (Grant No. 21340104).

\end{document}